\documentclass[preprint,aps]{revtex4}
\begin{document}

\newcommand{\be}{\begin{equation}}
\newcommand{\ee}{\end{equation}}
\newcommand{\ben}{\begin{eqnarray}}
\newcommand{\een}{\end{eqnarray}}
\newcommand{\n}{\nonumber  }
\newcommand{\nn}{\nonumber \\ }
\newcommand{\nd}{\noindent}
\newcommand{\p}{\partial}

\title{Legendre-transform structure derived from quantum theorems}

\author{S.P. Flego$^2$}

\author{A. Plastino$^{1,\,3}$}
\author{A.R. Plastino$^{3,\,4}$}

\affiliation{ $^{1}$Universidad Nacional de La Plata, Instituto de
F\'{\i}sica (IFLP-CCT-CONICET), C.C. 727, 1900 La Plata, Argentina
and
\\ Universitat de les Illes Balears and IFISC-CSIC,
  07122 Palma de Mallorca, Spain \\$^{2}$Universidad Nacional de La Plata, Facultad de
Ingenier\'{\i}a,
Depto. de Ciencias B\'asicas,  1900 La Plata, Argentina \\  $^{3}$ CREG-University of La Plata-CONICET\\
$^{4}$Instituto Carlos I de Fisica Teorica y Computacional and
Departamento de Fisica Atomica, Molecular y Nuclear, Universidad
de Granada, Granada, Spain }
%

\begin{abstract}
\nd By recourse to i) the Hellmann-Feynman theorem and ii)
the Virial one, the information-optimizing principle based on
Fisher's information measure uncovers a Legendre-transform
structure associated with Schr\"odinger's equation, in close
analogy with the structure that lies behind the standard
thermodynamical formalism. The present developments provide
new evidence for the information theoretical links based on
Fisher's measure that exist between Schr\"odinger's equation,
on the one hand, and thermodynamics/thermostatistics on the
other one.

\vspace{1.cm}


\end{abstract}

\maketitle

\section{Introduction}

\nd The thermodynamical formalism is characterized by its Legendre
transform structure \cite{deslog,callen}. Legendre transformations
allow us to express fundamental thermal equations in terms of a
set of independent variables chosen to be convenient for a given
problem \cite{deslog,callen}. In a more general context, Legendre
transform structures arise naturally in physical theories or models
that are based upon entropic or information theoretical optimization
principles. An example is that of references
\cite{evapla1,evapla2,evapla4,evapla8}, that purports to rederive
on such a basis the principles of statistical mechanics. Here we
will explore Schr\"odinger's non-relativistic equation for links
with the Legendre transform framework.

\nd The background for the present considerations was provided by
Jaynes, who established in the 50's a perdurable link between
Information Theory, Thermodynamics, and Statistical Mechanics
\cite{jaynes,katz}. At its core one finds a variational technique
 involving extremization of Shannon's logarithmic information measure
$$S= -\sum_k p_k \ln{p_k},$$
subject to  constraints imposed by the {\em a priori} knowledge at
hand concerning the system of interest. By identifying Shannon's
measure with the thermodynamic entropy a new foundation for statistical
mechanics was thereby obtained based on a general statistical inference
prescription \cite{jaynes,katz}. The ensuing methodology is usually known
as the MaxEnt-one \cite{jaynes,katz}. The MaxEnt approach provides an
insightful interpretation of the role played by the Legendre
transform referred to above, that is summarized in the Appendix.

\nd An approach similar to the one advanced by Jaynes
 was successfully developed many years later replacing
Shannon's $S$ above by Fisher's information measure (FIM)
 $I$  \cite{frieden,frieden2,frieden3,pla7} (see Section III below).
 Such developments  provided an additional perspective
 within the so-called Wheeler's program of establishing a
foundation for the basic laws of physics based on concepts
from information theory \cite{wheeler}. Considerable effort has been
expended recently on exploring the physical implications of Fisher
information. Indeed, the Fisher Information measure has been
successfully applied to the study of several physical scenarios
(as a non-exhaustive set, see for instance
\cite{frieden3,flego,pla2,pla3,pla4,olivares,silver,flavia,
KSC10,FS09,U09,LADY08,SAA07,N07,N06}).

\nd Both thermal-information connections, Shannon's and Fisher's,
are made by means of  a set if first-derivative relations (the
Legendre structure) that involve i) the Lagrange multipliers that
emerge from the variational process, ii) the information
quantifier ($S$ or $I$), and iii) the expectation values that
constitute the input, a-priori information on the system of
interest. In the Fisher's case a Schr\"odinger-like equation is
involved, a fact of paramount importance for our present purposes.

\nd We will show that the  virial and Hellmann-Feynman theorems,
essential quantum features, straightforwardly lead to a
Legendre-transform structure. After a preliminary presentation in
Section II we recapitulates the essential ingredients of Fisher's
thermodynamics \cite{flavia} in Section III. Our main result are
derived in Section IV.


\section{Preliminaries}

\subsection{Virial theorem}

\nd For any quantum system in stationary state, with a Hamiltonian does not involve time explicitly,
\ben \label{virial-1}
    H=~-~ \frac{\hbar^2}{2m}~\vec{\nabla}~+~U(\vec{x})~
\een the virial theorem states  that \cite{greiner}

\ben \label{virial-2} \left\langle ~ -~
\frac{\hbar^2}{m}~\vec{\nabla}~ \right\rangle = \left\langle
\vec{x}.\vec{\nabla} U(\vec{x})\right\rangle \een where the
expectation value is taken for stationary states of the
Hamiltonian.

\subsection{Feynman-Hellmann theorem}

\nd The Feynman-Hellmann theorem (HFT) \cite{hf1,hf2,hf3,hf4,hf5}
establishes  the relationship between perturbations in an operator
on a complex inner product space and the corresponding
perturbations in the operator's eigenvalues. It shows that to
compute the derivative of an eigenvalue with respect to a
parameter of the operator we need only know the corresponding
 eigenvector and the derivative of the operator. More to the point,
the Hellmann-Feynman (HF) theorem refers to a parameter dependent
eigen-system. It asserts that, in the case of a hermitian operator
$H(\lambda)$ (whose eigenvectors are $\psi_i$),  a non-degenerate
eigenvalue $E_i$ varies with respect to the parameter $\lambda$
according to the expression

\ben  \label{HF-l} \frac{\partial E_i}{\partial \lambda}= \langle
\psi_i \vert \frac{\partial H}{\partial \lambda}  \vert \psi_i
\rangle. \een \nd The theorem has a rich history and many
applications, that can be consulted in \cite{hf2}. The FH theorem
can be proved to hold for exact eigenstates and also for
variationally determined states \cite{hf3}.

\section{The Fisher thermal formalism}
\subsection{Basics results}
\nd This formalism was advanced in  Ref. \cite{pla7}. One
considers a system that is specified by a physical parameter
$\theta$, while the quantity $f(\theta)$ determines the normalized
probability distribution function (PDF) for it. Fisher's
Information Measure (FIM) $I$ gets defined as \be \label{eq.1-1} I
\,=\,\int ~dx ~f(x,\theta) \left\{\frac{\partial ~ }{\partial
\theta}~\ln{[f(x,\theta)]}\right\}^2.\ee Fix attention upon
 translational families, which are mono-parametric distribution ones of
the form
    \[f(x,\theta)=f(x-\theta),\]
 known up to the shift parameter $\theta$. All members of the family possess
identical shape, and for them  FIM adopts the simpler form \be
\label{eq.1-2} I \,=\,\int ~dx ~f(x)\left\{\frac{\partial ~
}{\partial x}~\ln{[f(x)]}\right\}^2.\ee


\nd We are interested in a system that is specified by a set of
$M$ physical parameters $\mu_k$. More to the point \ben
\label{eq.1-3} \mu_k = \langle A_{k}\rangle\hspace{0.2cm},
 \hspace{0.5cm} & {\rm with}&\hspace{0.2cm} A_{k}=
 A_{k}(x)\hspace{0.3cm}(k=1,...,M).
\n\een The set of $\mu_{k}$-values   represents  the empirical
information at hand. If the pertinent probability distribution
function (PDF)  is $f(x)$, then

\be \label{eq.1-4} \langle A_{k}\rangle\,=\,\int ~dx ~A_{k}(x)
~f(x), \hspace{0.5cm} k=1,\dots ,M. \ee These mean values will
play the role of extensive thermodynamical variables
 \cite{pla7}. The relevant PDF $f(x)$ for us is the one that extremizes
 (\ref{eq.1-2}) subject to i) the prior conditions (\ref{eq.1-3})
and, of course, ii) the normalization condition \be \label{eq.1-5}
\int ~dx ~f(x) ~= ~1. \ee

\nd Accordingly, the extremization problem that we face is \be
\label{eq.1-6}\delta \left( I - \alpha \int ~dx ~f(x) -
\sum_{k=1}^M~\lambda_k\int ~dx ~A_{k}(x) ~f(x)\right) = ~0 \ee
where we have  $(M+1)$ Lagrange multiplier. Variation leads to \be
\label{eq.1-7} \left[\frac{1}{f^{2}}~\left(\frac{\partial
f}{\partial x}\right)^2 +~\frac{\partial~}{\partial x}\left(
\frac{2}{f}~ \frac{\partial f}{\partial x}\right)\right] + \alpha
+ \sum_{k=1}^M~\lambda_k~A_k(x) = ~0 \ee It is convenient
\cite{silver,pla7,paul} to  introduce a function $\psi(x)$ via the
identification $\psi(x)^2 =f(x)$ so that Eq. (\ref{eq.1-7})
acquires a wave equation form (SWE)

 \be \label{eq.1-8} -~\frac{1}{2}~\frac{\partial^2 ~}{\partial x^2} \psi~-~\sum_{k=1}^{M}~
 \frac{\lambda_{k}}{8}~ A_{k}\,\psi ~= ~ \frac{\alpha}{8}~ \psi,
 \ee which can be formally interpreted as a Schr\"odinger equation
 for a particle of unit mass moving in the effective, ``information"
 pseudo-potential [Cf. Eq. (\ref{eq.1-4})]

\be \label{eq.1-9} U~=~U(x)
=~-\frac{1}{8}~\sum_{k=1}^{M}\,\lambda_{k}~ A_{k}(x). \ee

\nd The Lagrange multiplier ($\alpha /8$) plays the role of an
energy eigenvalue $E=\alpha/8$. The Lagrange parameters
$\lambda_k$ are fixed, of course,  by recourse to the input prior
information.
 Notice that the eigen-energies $\alpha/8$ yield automatically the value of
 the Lagrange multiplier associated to normalization. The  square of the solutions $\psi$ is the desired PDF

  \be
\label{eq.1-10} \psi(x)^2 ~= ~ f(x). \ee

\subsection{Finding a convenient way of using FIM}

\nd It is now important to establish a new form of expressing
Fisher's information measure as a function of $\psi$. One
substitutes ({\ref{eq.1-10}}) into Eq.  ({\ref{eq.1-2}}) to find

\ben \label{eq.1-11} I \, = \,\int dx ~f \left(\frac{\partial
\ln{f} }{\partial x}\right)^2\,= \, \int dx ~ \psi_n^2 ~
\left(\frac{\partial \ln{\psi_n^2} }{\partial x} \right)^2\,=\, 4
\int d x ~ \left(\frac{\partial \psi_n }{\partial x} \right)^2
\een which can be re-expressed as \ben \label{eq.1-12} I \, =\,
-~4 \int \psi_n \frac{\partial^2 ~}{\partial x^2} \psi_n~dx= - 4
\left\langle\frac{\partial^2 ~}{\partial x^2}\right\rangle \een
Now, using the SWE (\ref{eq.1-8}) one obtains

 \ben \label{eq.1-13}
I\,=\, \int ~ \psi_n \left(\alpha +
\sum_{k=1}^M~\lambda_k~A_k\right) \psi_n~dx. \een Finally, the
prior conditions (\ref{eq.1-3}) and the normalization condition
(\ref{eq.1-5}) allow one  to express $I$ in the quite convenient
fashion

\ben \label{eq.1-14} I(\left\langle
A_1\right\rangle,\ldots,\left\langle A_M\right\rangle) \,=\,\alpha
 + \sum_{k=1}^M~\lambda_k\left\langle A_k\right\rangle. \een


\subsection{Fisher thermodynamics}

 \nd    The connection between the
  solutions of   Eq. (\ref{eq.1-8})  and thermodynamics
  has been established in Refs. \cite{pla7} and \cite{flego}.
 We summarize now the main details. The reciprocity relations (\ref{katz}) and
 their Fisher-counterparts  are an expression of the
Legendre-transform structure of thermodynamics \cite{PP,flavia}
and constitute its essential formal ingredient \cite{deslog}. It
is of the essence that they also hold for the Fisher treatment.
Standard thermodynamic makes use of the derivatives of the entropy
$S$ with respect to both the $\lambda_{i}$ and $\langle
A_{i}\rangle$ quantities (for instance, pressure and volume,
respectively).

\nd Analogous properties of $\partial I/\partial \lambda_{i}$ and
$\partial I/\partial \langle A_{i}\rangle$ are valid as well
\cite{pla7}. We start with (\ref{eq.1-14}) and consider its
Legendre transform, that we call $\alpha$, i.e.,

 \ben \label{eq.R-4}
\alpha(\lambda_1,\ldots,\lambda_M)= I(\left\langle
A_1\right\rangle,\ldots,\left\langle A_M\right\rangle) -
\sum_{k=1}^M~\lambda_k\left\langle A_k\right\rangle , \een so that
\be \label{eq.R-5}\frac{\partial \alpha}{\partial \lambda_{i}}= -
\langle A_i\rangle,  \ee

\nd and recall two expressions obtained in \cite{pla7}, namely,

\be \label{eq.R-6} \frac{\partial I }{\partial \left\langle A_k
\right\rangle}\,=\,\lambda_k , \ee and
\ben \label{eq.R-7} \frac{\partial I}{\partial
\lambda_{i}}=\sum_{k}^{M} \lambda_{k} \frac{\partial \langle
A_{k}\rangle}{\partial \lambda_{i}} \een

\nd which is a generalized Fisher-Euler theorem. It is instructive
to glance at the Appendix at this point to note that entirely
similar relations are obeyed by the ordinary Gibbs-Boltzmann
entropy $S$. On the basis of such an observation, it seems natural
to consider that the three reciprocity relations above should
allow one to speak of a ``Fisher-thermodynamics" \cite{flego}.
Curiously enough, it can be shown that the HF theorem can be
looked at as a reciprocity relation of the type (\ref{eq.R-5})
\cite{nuestro1}.


\section{A quantal-Fisher  connection}

\nd We begin here to develop the original contents of this
presentation. Let us consider now that Eq. (\ref{eq.1-8}) is an
ordinary  Schr\"odinger wave equation for a particle of unit mass
in which the Lagrange multiplier ($\alpha /8$) plays the role of
an energy eigenvalue $E=\alpha/8$. Remark that
 $U(x)$ is taken now to be an actual, physical potential, not an
 effective, ``information" one.
This is the starting point. \vskip 2mm

\nd  We emphasize now the fact that our FIM $I$ is now seen
to be proportional to the expectation value of the Laplace operator,
namely,

 \ben \label{virial-3}
 I \, = \,\int dx ~f \left(\frac{\partial \ln{f} }{\partial
x}\right)^2\,= \,-~4 \int \psi_n \frac{\partial^2 ~}{\partial x^2}
\psi_n~dx ~=~- 4~\left\langle \frac{\partial^2 ~}{\partial x^2}
\right\rangle, \een \nd where $\psi_n$ are the eigenfunctions of

\be \label{virial-4}
\left[-~\frac{1}{2}~\frac{\partial^2~}{\partial x^2} -~
\frac{1}{8} \sum_k \,\lambda_k\, A_k~\right]\psi ~= ~
\frac{1}{8}~\alpha~ \psi.\ee

\nd We take it for granted that, as customary, the potential $U$
admits of a series-expansion (powers of $x^k$) of the form

 \be \label{virial-5} U(x)= -~ \frac{1}{8} \sum_k \,
 \lambda_k\, A_k   \equiv -~ \frac{1}{8} \sum_k \,\lambda_k\,x^k=
 \sum_k\, a_k
 x^k;\,\,\,\,\,-\lambda_k/8=a_k. \ee Thus, the $A_k$ in the preceding Sections become here $x^k-$moments and one assumes that the expansion
 is good enough if $M$ terms of them are included. The $\lambda_k$
 are now the expansion-coefficients and not Lagrange multipliers.
 A Fisher's measure is to be constructed with these coefficients.
 Recourse to  the Virial theorem (\ref{virial-2}) allows us to cast the
 FIM-expression (\ref{virial-3}) in the fashion
 \ben \label{virial-7} I \, =\, -~
~\sum_{k=1}^{M}\,\frac{k}{2}~\,\lambda_{k}~\left\langle
A_{k}\right\rangle. \een

\nd Now, replacing (\ref{virial-7}) into (\ref{eq.1-14}) and
solving for $\alpha$ one finds

 \ben \label{virial-8} \alpha \,=\, -
\sum_{k=1}^M~ \left(\frac{k}{2}+1\right)\lambda_k\left\langle
A_k\right\rangle, \een having thus obtained two expressions that
pave a direct road towards our present goal.

\subsection{Hellmann-Feynman and Virial theorems imply reciprocity
relations}

\nd  In this subsection we are going to show that  Eqs.
(\ref{virial-7}), (\ref{virial-8}), and the Hellmann-Feynman
theorem (\ref{HF-l}) jointly lead to Fisher-reciprocity relations.
It being clear up this point that the $\lambda$'s are
expansion-coefficients, we will speak herefrom only
``$\lambda$-language".

\vspace{0.5cm} \nd In one dimensional scenarios, the
eigenfunctions $\psi(x)$  of (\ref{eq.1-8}) are real. We appeal
now to the {\it Hellmann-Feynman theorem} and  obtain

\be \label{HFV-RR-1} \frac{\partial }{\partial
\lambda_k}\left(\frac{\alpha}{8}\right)~=~ \langle \psi \vert
\frac{\partial H}{\partial \lambda_k}  \vert \psi \rangle =
\langle \psi \vert -~\frac{1}{8}~A_k  \vert \psi \rangle
\hspace{1.cm} \longrightarrow \hspace{1.cm} \frac{\partial
\alpha}{\partial \lambda_{k}} = - \langle A_k \rangle, \ee thus
discovering that the HF theorem immediately implies  the
reciprocity relation (\ref{eq.R-5}).

\vspace{0.3cm}




\nd It is clear that differentiating (\ref{virial-8}) with respect
to $\lambda_{n}$ yields

\ben \label{HFV-RR-3}
 \frac{\partial \alpha}{\partial \lambda_{n}}= -\left(\frac{n}{2}+1\right) \left\langle
A_n\right\rangle-\sum_{k=1}^M~\left(\frac{k}{2}+1\right)\lambda_k~
\frac{\partial \left\langle A_k\right\rangle}{\partial
\lambda_{n}}. \een

 \nd The two relations (\ref{HFV-RR-1}) and (\ref{HFV-RR-3}) result now in

  \ben
\label{HFV-RR-4}
 \frac{n}{2}~\left\langle A_n\right\rangle = -
\sum_{k=1}^M~\left(\frac{k}{2}+1\right)\lambda_k~\frac{\partial
\left\langle A_k\right\rangle}{\partial \lambda_{n}}. \een

\nd We go back to (\ref{virial-7}) at this point and differentiate
it with respect to
 $\lambda_{n}$ to arrive at

  \ben \label{HFV-RR-5}
 \frac{\partial I}{\partial \lambda_{n}}= -\frac{n}{2}~\left\langle
A_n\right\rangle-\sum_{k=1}^M~\frac{k}{2}\lambda_k~ \frac{\partial
\left\langle A_k\right\rangle}{\partial \lambda_{n}}. \een

\nd At this stage, recourse to  the relation (\ref{HFV-RR-4})
allows one to recover the Euler relations

\ben \label{HFV-RR-6}
 \frac{\partial I}{\partial \lambda_{n}}= \sum_{k=1}^M~\lambda_k~
\frac{\partial \left\langle A_k\right\rangle}{\partial
\lambda_{n}}. \een

\nd We also have

 \ben \label{HFV-RR-7}
 \frac{\partial I(<A_1>,\ldots,<A_M>)}{\partial \lambda_{n}}= \sum_{k=1}^M~\frac{\partial I}{\partial \left\langle A_k\right\rangle}~
\frac{\partial \left\langle A_k\right\rangle}{\partial
\lambda_{n}}, \een so that, comparing (\ref{HFV-RR-6}) and
(\ref{HFV-RR-7}) we immediately obtain

\ben \label{HFV-RR-8} \frac{\partial I}{\partial \langle A_n
\rangle}= \lambda_{n}.
 \een

\nd The three expressions (\ref{HFV-RR-1}), (\ref{HFV-RR-6}) and
(\ref{HFV-RR-8}), obtained by application of the Hellmann-Feynman
theorem and  the Virial one to Fisher's information measure, are
reciprocity relations that in turn constitute a  manifestation of
 an underlying Legendre-invariant structure, analogous to that of thermodynamics, our main result
 here.

\section{Conclusions}

\nd In this work we have shown that, if Fisher's measure $I$ is
associated to a Schr\"odinger wave equation (SWE), as it happens
whenever one extremizes it  subject to appropriate constraints,
two theorems intimately linked to the SWE, the Hellmann-Feynman
and Virial ones, automatically  lead to  Jaynes-like reciprocity
relations involving the coefficients of the  series-expansion of
the potential function. One may then dare to assert that a
Legendre-transform structure seems to underly the one-dimensional
non-relativistic Schr\"odinger's equation, a rather surprising
finding.

\noindent
{\bf Acknowledgments-} This work was partially
supported by the Project FQM-2445 of the
Junta de Andalucia, Spain.

\newpage

\section{Appendix: MaxEnt and Legendre structure}

The classical MaxEnt probability distribution function (PDF),
associated to Boltzmann-Gibbs-Shannon's logarithmic entropy, is
given by \cite{jaynes,katz}

\be f(MaxEnt)=f(x)=\exp \left\{-\left[
\Omega+\sum_{i=1}^M\,\lambda_i\,A_i(x) \right]\right\}, \ee

\noindent with \cite{jaynes,katz}

\be \Omega(\lambda_1,\ldots,\lambda_M)=\ln \left\{ \int dx\,
\left[ \exp\left(-\sum_{i=1}^M\,\lambda_i\,A_i(x)\right) \right]
\right\} \equiv -\lambda_o, \label{omega}\ee

\be \frac{\partial \Omega(\lambda_1,\ldots,\lambda_M)}{\partial
\lambda_j}=   -\langle A_j\rangle, \,\,\,\,\,\,\,\,(j=1,\ldots,M),
\ee and

\be S=\Omega+\sum_{i=1}^M\,\lambda_i\,\langle A_i \rangle,
\label{expandS} \ee entailing \be  dS=
\sum_{i=1}^M\,\lambda_i\,d\langle A_i\rangle. \ee

\noindent The  Euler theorem holds \cite{katz} \be \frac{\p S}{\p
\lambda_i}=\sum_k \lambda_k \frac{\p \langle A_k \rangle}{\p
\lambda_i},\label{euler}\ee and, using (\ref{expandS}), one
arrives to

\ben dS
&=&\sum_{i=1}^M\,\lambda_i\,d\langle
A_i\rangle\,\,\,\Longrightarrow \frac{\partial S}{\partial \langle
A_i\rangle }=\lambda_i\nn S&=& S(\langle A_1\rangle,\ldots,\langle
A_M\rangle). \een Effecting now  the Legendre transform \be \Omega
=\Omega(\lambda_1,\ldots,\lambda_M)=S-\sum_{i=1}^M\,\lambda_i\,\langle
A_i\rangle, \ee one immediately ascertains that reciprocity holds,
namely,

\be  \frac{\partial S}{\partial \langle A_j\rangle}
=\lambda_j\,\,\,\,\,\,\, {\rm and} \,\,\,\,\, \frac{\partial
\Omega}{\partial \lambda_j }=-\langle A_j\rangle;\,\,\,
j=1,\ldots,M, \label{katz} \ee where the second set of equations,
 together with (\ref{omega}), yield the Lagrange multipliers as a
function of the input information regarding expectation values
\cite{katz}. The reciprocity relations (\ref{euler}) +
(\ref{katz}) are a manifestation of the Legendre-transform
structure of thermodynamics \cite{deslog,PP} and its most salient
structural mathematical feature.


\begin{thebibliography}{99}


\bibitem{deslog} Deslog E. A., {\em Thermal Physics} (Holt,
Rinehart and Winston, New York, 1968).

\bibitem{callen} H. B. Callen, {\it Thermodynamics} (Wiley, NY, 1960).



\bibitem{evapla1} E. Curado, A. Plastino, Phys. Rev. E {\bf 72} (2005) 047103.

\bibitem{evapla2}  A. Plastino, E. Curado, Physica A {\bf 365} (2006)
24.


\bibitem{evapla4}  A. Plastino, E. Curado, Physica A {\bf 386}  (2007) 155.








\bibitem{evapla8}
E. Curado, F. Nobre, A. Plastino, Physica A {\bf 389} (2010) 970.



\bibitem{jaynes} Jaynes E.T., Information Theory and Statistical Mechanics, Phys. Rev.
 {\bf 1957} {\em 106}, 620-630.

\bibitem{katz} Katz, A.~~ {\it Principles of Statistical Mechanics:
The Information Theory Approach} (Freeman and Co.: San Francisco,
1967).

\bibitem{PP} A. Plastino, A. R. Plastino,  Phys. Lett. A {\bf 226}    (l997) 257.

\bibitem{frieden} B. R. Frieden, Phys.  Rev. A {\bf 41} (1990) 4265.

\bibitem{frieden2}  B. R. Frieden, {\em Physics from Fisher information measure}
(Cambridge, University Press; Cambridge, 1998).

\bibitem{frieden3}  B. R. Frieden, B. H. Soffer, Phys. Rev. E {\bf 52 1995} (1995) 2274.

\bibitem{pla7}  B. R. Frieden, A. Plastino, A. R. Plastino, B. H. Soffer,
 Phys.  Rev. E  {\bf 60 1999} (1999) 48.

\bibitem{wheeler} Wheeler J.A.,  {\em Complexity, entropy and the physics of information};   (W.H. Zurek Ed.; Addison Wesley; New York, 1991, pp. 3-28).

\bibitem{flego} S. P. Flego, B. R. Frieden, A. Plastino, A. R. Plastino, B. H.  Soffer,  Phys. Revi. E {\bf 68} (2003) 016105.

\bibitem{pla2} A. R. Plastino, A. Plastino, Phys. Rev.  E {\bf 54} (1996) 4423.

\bibitem{pla3} A. Plastino, A. R. Plastino, H. G. Miller, Phys. Lett. A
 {\bf 235} (1997) 129.

\bibitem{pla4} A. R. Plastino, M. casas, A. Plastino, Phys.  Lett. A {\bf 246} (1998) 498.

\bibitem{olivares} F. Olivares, F. Pennini, A. Plastino,
 Physica A {\bf 389} (2010) 2218.

\bibitem{silver}  Silver R. N. in {\em E. T. Jaynes: Physics and Probability}
(W. T. Grandy and P. W. Milonni, Eds.; Cambridge University Press;
Cambridge, 1992).

\bibitem{flavia} F. Pennini, A. Plastino, Phys. Rev. E {\bf 71} (2005) 047102.


\bibitem{KSC10} V. Kapsa, L. Skala and J. Chen,
Physica E {\bf 42} (2010) 293.

\bibitem{FS09} B.R. Frieden and B.H. Soffer,
Physica A {\bf 388} (2009) 1315.

\bibitem{U09} M. R. Ubriaco. Phys. Lett. A
{\bf 373} (2009) 4017.

\bibitem{LADY08} S. Lopez-Rosa, J.C. Angulo, J.S. Dehesa
and R.J. Yanez, Physica A {\bf 387} (2008) 2243.

\bibitem{SAA07} K.D. Sen, J. Antolin and J.C. Angulo,
Phys. Rev. A {\bf 76} (2007)  032502.

\bibitem{N07} A. Nagy, Chem. Phys. Lett. {\bf 449}
(2007) 212.

\bibitem{N06} A. Nagy, Chem. Phys. Lett. {\bf 425} (2006) 154.


\bibitem{greiner} W. Geiner and B. M\"uller, {\it Quantum mechanics. An
Introduction}(Springer, Berlin, 1988).

\bibitem{hf3}  W. Namgung,  Journal of the Korean Physical Society, {\bf 32} (1998) 647.

\bibitem{hf4}H. G. A. Hellmann, Z.  Phys, {\bf 85}  (1933) 180.

\bibitem{hf5} R. P. Feynman, Phys. Rev. {\bf 56} (1939)  340.

\bibitem{hf1} D. J. Griffiths, {\it Introduction to Quantum Mechanics} (Prentice Hall, Englewood Cliffs, NJ,
1995).

\bibitem{hf2}  David W- Wallace, {\it An introduction to Hellmann-Feynman
theory}(Master Thesis, University of Central Florida, Orlando,
Florida, 2005 (unpublished)).

\bibitem{nuestro1} S. P. Flego, A. Plastino, A. R. Plastino,
 unpublished.

\bibitem{paul}  Richards P. I., {\em Manual of mathematical physics }
(Pergamon Press; London,  1959).


\end{thebibliography}

\end{document}